\documentclass[12pt]{article}%
\usepackage[a4paper,margin=1in]{geometry}
\usepackage{amsmath,amssymb}
\usepackage{graphicx}
\usepackage{cite}
\usepackage{amsmath}
\usepackage{amsfonts}
\usepackage{amssymb}%
\usepackage{array}
\usepackage{booktabs}
\usepackage{ragged2e}
\usepackage{float}
\providecommand{\U}[1]{\protect\rule{.1in}{.1in}}
\begin{document}

\title{Probing Lorentz Invariance Violation at High-Energy Colliders via Intermediate
Massive Boson Mass Measurements: Z Boson Example}
\author{Z. Kepuladze$^{1}$, J.~Jejelava$^{1}$\\{\small $^{1}$Andronikashvili Institute of Physics, Tbilisi State University,
Ilia State University}}
\date{}
\maketitle

\begin{abstract}
Lorentz invariance (LI) is a foundational principle of modern physics, yet its
possible violation (LIV) remains an intriguing window to physics beyond the
Standard Model. While stringent constraints exist in the electromagnetic and
hadronic sectors, the weak sector---particularly unstable bosons---remains
largely unexplored. In this work, based on our recent studies and conference
presentation, we analyze how LIV manifests in high-energy collider
experiments, focusing on modifications of Z boson dispersion relations and
their impact on resonance measurements in Drell--Yan processes. We argue that
precision measurements of resonance masses at colliders provide sensitivity to
LIV at the level of $10^{-9}$, comparable to bounds derived from cosmic rays.
We also discuss the interplay between LIV and gauge invariance, highlighting
why only specific operators provide physical effects. The phenomenological
implications for both Z and W bosons are outlined, with emphasis on
experimental strategies for current and future colliders.

\end{abstract}

\section{Introduction}

Lorentz invariance (LI) underpins the structure of quantum field theory and
general relativity. Yet possible violations of this symmetry are theoretically
motivated. Spontaneous Lorentz symmetry breaking can give rise to emergent
vector or tensor degrees of freedom, which may act as Goldstone modes of
spacetime symmetry violation \cite{SLIV}. Alternatively, LIV can appear as a
mechanism for ultraviolet completion, rendering theories finite, as in
Ho\v{r}ava's construction \cite{Horava}. At low energies, if present, LIV
should be strongly suppressed, but in the ultra--high-energy domain it may
have significant phenomenological consequences.

The motivation for studying LIV partly comes from cosmic ray physics. The
famous Greisen--Zatsepin--Kuzmin (GZK) cutoff \cite{GZK} predicts that
ultra--high-energy cosmic rays above $\sim5\times10^{19}$ eV should be
strongly attenuated by interactions with the cosmic microwave background.
However, experiments such as AGASA and, later, the Pierre Auger Observatory
reported excess events beyond the GZK bound, sparking interest in LIV as a
possible explanation \cite{AGASA, Auger}. Such cosmic ray observations thus
continue to provide motivation for precise LIV studies.

Another motivation comes from neutrino physics. The possibility of neutrinos
traveling at speeds slightly different from light has been considered in
multiple contexts. Early hints, such as the controversial OPERA result,
suggested superluminal neutrinos, though this was later shown to be an
experimental error. Nonetheless, neutrino time-of-flight experiments and
supernova neutrino observations (e.g., SN1987A) constrain deviations from the
speed of light at the level of $10^{-9}$ or better \cite{Neutrino}. These are
directly relevant because they probe LIV in the weak sector, where constraints
remain far weaker than in the electromagnetic one.

Thus, cosmic ray and neutrino data both highlight the importance of searching
for LIV in controlled environments such as high-energy colliders. While
astrophysical observations give very stringent limits in some channels,
colliders allow systematic and model-independent tests, especially for
unstable weak bosons that cannot be probed astrophysically.

\section{Constraints and Open Windows}

The tightest constraints on LIV arise mostly from astrophysical processes,
since cosmic rays may carry energies far beyond those accessible at
accelerators or other Earth-based experiments. If we describe these
restrictions in the language of possible deviations from the maximal
attainable velocity for a given particle species (a particle's
\textquotedblleft speed of light,\textquotedblright\ so to speak)
\cite{Colleman-Gleshow}, defined as $\delta=\Delta c/c$, then the following
constraints can be quoted. For electrons: $\left\vert \delta_{e}\right\vert
<10^{-19}$ \cite{Kostelecky-Russell, Alt}, $\left\vert \delta_{e}%
-\delta_{\gamma}\right\vert <5\cdot10^{-19}$ \cite{Photon}. Restrictions on
photons and protons fall in approximately the same range
\cite{Kostelecky-Russell}. These arise from the non-observation of otherwise
expected effects in the presence of LIV, such as photon decay and vacuum
Cherenkov radiation (generally derived from threshold-energy arguments), tests
of rotating optical cavities, vacuum birefringence, dispersion,
Michelson--Morley--type resonators, or time-of-flight measurements. Bounds
from these processes are so strong that they effectively rule out LIV in the
QED sector at accessible scales.

Neutrinos provide a different picture. Time-of-flight measurements constrain
their velocity relative to light. Supernova SN1987A neutrino arrival times
imply $\left\vert \delta_{\nu}\right\vert <10^{-9}$ at energies of tens of
MeV. At higher energies, IceCube measurements of PeV neutrinos place bounds at
the level of $10^{-10}$--$10^{-11}$. Atmospheric neutrino oscillations
observed by Super-Kamiokande constrain certain LIV coefficients to $10^{-8}$
in the GeV range.

In general, unstable particles---among them the weak bosons W and Z, which are
both unstable and short-lived---evade such astrophysical probes. Their
dispersion relations have never been directly tested outside collider
environments. Consequently, the weak sector remains essentially unconstrained
with respect to LIV. Whether this is a special feature of the weak sector
remains to be determined. At the same time, this gap represents an open
experimental window: accelerator experiments can probe parameter space that
astrophysical methods cannot access. Collider studies of massive intermediate
boson resonances therefore allow us to test LIV systematically in a sector
that has remained hidden from astrophysical scrutiny.

The present contribution builds upon our recent studies \cite{KEP,KEP-JEJ}.

\section{Testing LIV at Accelerators: Concept}

Whatever the origin of LIV might be, the low-energy phenomenology can always
be parameterized by possible modifications to particle propagation and
interactions. If the preferred direction of LIV is fixed in spacetime by a
timelike or spacelike unit vector $n_{\mu}=(n_{0},\overrightarrow{n})$
\footnote{The preferred direction, fixed by the $n_{\mu}$ vector, transforms
as a constant four-vector under Lorentz transformations. The explicit form
written in the text corresponds to a particular reference frame; in other
frames the components of $n_{\mu}$ change accordingly, while its invariant
character as a preferred direction remains.}, one of the simplest
renormalizable interactions between the vector field $A_{\mu}$ and fermion
field $\Psi$ may take the form
\begin{equation}
e\delta_{\text{int}}(A_{\mu}n^{\mu})\overline{\Psi}(\gamma_{\nu}n^{\nu})\Psi
\end{equation}

To detect such modifications at accelerators, one usually examines their
effect on cross sections, which acquire the general form
\begin{equation}
\sigma_{LIV}=\sigma_{LI}(1+\delta_{\text{int}}f(\Omega,n))
\end{equation}
where $f(\Omega,n)$ encodes the interplay between the preferred LIV direction
$n_{\mu}$ and the orientation of the process in spacetime. Because $n_{\mu}$
picks out a special direction, it introduces anisotropy into the process.
Since an accelerator rotates with the Earth, the relative orientation with
respect to $n_{\mu}$ also changes with sidereal time. Therefore, daily
modulations should emerge in the cross section if LIV is present and the
experimental accuracy is sufficient.

Searches for such modulations have been conducted at the Large Hadron Collider
(LHC), yielding limits of $\left\vert \delta_{\text{int}}\right\vert <10^{-5}$
\cite{CMS, Lunghi}. While the specific modifications studied in those analyses
originated in the quark sector, their functional form is quite general and can
be applied to a broad class of LIV operators, including the interaction above.
Provided that $f(\Omega,n)$ has no strong energy dependence, these bounds can
be generalized accordingly.

A noticeable property of this form of $\sigma_{LIV}$ is that LIV contributions
always enter at the same order in $\delta_{\text{int}}$. Detectability
therefore depends only on experimental precision, essentially independent of
the energy scale. This situation contrasts with modifications of the
dispersion relation, which in their simplest form can be written as
\begin{equation}
p^{\mu}p_{\mu}=M_{B,eff}^{2}=M_{B}^{2}+\delta\,E^{2} \label{M1}%
\end{equation}
Here $p_{\mu}=\left(  E,\overrightarrow{p}\right)  $ is the four-momentum and
$M_{B}$ the given boson mass. We have also introduced notion of the effective
mass. In this case the LIV term competes with the mass term, and the hierarchy
between them changes with energy, making LIV effects increasingly accessible
at high energies. Scattering processes mediated by a massive intermediate
boson therefore become highly sensitive to such modifications in the resonance region.

For the boson with (\ref{M1}), one can approximate \cite{SME}
\begin{equation}
\Gamma_{LIV}=\frac{M_{B,eff}^{2}}{M_{B}^{2}}\Gamma_{LI}%
\end{equation}
so the unstable boson propagator becomes
\begin{align}
D  &  =\dfrac{i}{p_{\alpha}^{2}-M_{B}^{2}}\rightarrow\frac{i}{p_{\alpha}%
^{2}-(M_{B,eff}-ip_{0}\Gamma_{LIV}/2M_{B,eff})^{2}}\nonumber\\
&  =\frac{i}{p_{\alpha}^{2}-M_{B,eff}^{2}(1-i\Gamma_{LI}/2M_{B})^{2}}%
\end{align}

Consequently, the cross section is%
\begin{equation}
\sigma_{B}^{_{LIV}}\sim\left\vert D_{B}\right\vert ^{2} \label{sigmaLIV}%
\end{equation}
and the resonance mass $M_{res}$ now measures the effective mass instead:%
\begin{equation}
M_{res}^{2}=M_{B,eff}^{2}(1-\Gamma_{LIV}^{2}/4M_{B}^{2})
\end{equation}

Applying this framework to the weak Z boson, and comparing the resonance mass
shift with the current precision of $M_{Z}$, one finds that at LHC energies of
$E=14$ TeV, the present experimental uncertainty of $\Delta M_{Z}=\left\vert
M_{Z}-M_{Zresonance}\right\vert \approx2$ MeV \cite{Zmass} implies
\begin{equation}
\left\vert \delta\right\vert \approx\frac{2M_{Z}\Delta M_{Z}}{E^{2}}%
\approx2\cdot10^{-9} \label{est1}%
\end{equation}
This level of sensitivity is comparable to astrophysical constraints for the neutrinos.

Such a preliminary result is already convincing enough to justify further
investigation. In the next section we turn to the realistic Drell--Yan cross
section mediated by the neutral weak boson. When considering Lorentz
invariance violation in the weak sector, the Z boson provides the cleanest
probe due to its narrow resonance and well-measured leptonic decay modes.

\section{Modified Dynamics of the Z Boson}

To introduce modified dynamics for the neutral weak boson, a natural starting
point is to modify the kinetic term of the Z-boson Lagrangian. For a preferred
direction fixed in spacetime by the vector $n_{\mu}$, the kinetic term that
introduces LIV, modifies the dispersion relation, and is constrained by two
derivatives (i.e. is renormalizable), is%
\begin{equation}
\Delta L_{LIV}=\frac{\delta_{LIV}}{2}(\partial_{n}Z^{\mu})(\partial_{n}Z_{\mu
}),\text{ \ \ \ \ \ }\partial_{n}\equiv n_{\mu}\partial^{\mu} \label{LIV-L}%
\end{equation}

Alongside this term, one can introduce additional LIV operators,%

\begin{equation}
\Delta L_{LIV}=\frac{\delta_{LIV}}{2}(\partial_{n}Z^{\mu})(\partial_{n}Z_{\mu
})+\frac{\delta_{1LIV}}{2}(\partial_{\mu}Z_{n})(\partial^{\mu}Z_{n}%
)+\delta_{2LIV}(\partial_{\mu}Z^{\mu})(\partial_{n}Z_{n})
\end{equation}
with $Z_{n}\equiv n_{\mu}Z^{\mu}$.

These operators are often introduced because in the literature there is a
frequent attempt to enforce a gauge-invariant (GI) form. In that case one sets
{\normalsize $\delta_{LIV}=\delta_{1LIV}=-\delta_{2LIV}$. } However, neither
of the two additional terms influences the dispersion relation. More
importantly, LIV and GI do not go hand in hand. One can safely claim that if
we want physical LIV in a theory, gauge invariance must be broken at least
slightly. In fact, it is possible to obtain GI precisely from the demand that
LIV be physically unobservable \cite{SLIV, gauge}.

A simple demonstration is as follows. If we introduce a mass term
$m^{2}(n_{\mu}A^{\mu})^{2}$ (or any operator for that meter of the form
$F(n_{\mu}A^{\mu})$, $F(n_{\mu}A^{\mu})\overline{\Psi}\Psi$, $F(n_{\mu}A^{\mu
})\overline{\Psi}n_{\lambda}\gamma^{\lambda}\Psi$, etc.) into an otherwise GI
theory, then by performing a gauge transformation toward the axial gauge
$n_{\mu}A^{\mu}=0$ we can effectively eliminate LIV from the theory. Instead
of genuine LIV, we only succeed in fixing a particular gauge. The same is true
for any $F(A^{\mu})$. If the gauge equation $F(A^{\mu}+\partial^{\mu}%
\omega)=0$ has a solution for $\omega$ for arbitrary $A^{\mu}$, then
$F(A^{\mu})=0$ will become simply a gauge choice. If this gauge equation does
not have a solution, gauge invariance is broken and physical LIV necessarily manifests.

A distinct case arises if $F(A^{\mu})$ has a manifestly GI form itself, for
example $F(A^{\mu})\sim\delta n_{\mu}F^{\mu\lambda}n^{\nu}F_{\nu\lambda}$,
where $F_{\nu\lambda}=\partial_{\nu}A_{\lambda}-\partial_{\lambda}A_{\nu}$ and
$\delta$ is the LIV strength. In such a scenario, GI remains exact and gauge
can be fixed by our convenience. At first glance everything appears
consistent: the massless vector field still describes two propagating degrees
of freedom and the Coulomb law is intact. But once U(1) symmetry is broken, a
problem emerges: the vector field still carries only two degrees of freedom
instead of behaving as a massive vector should. In other words, the theory
becomes inconsistent with reality.

Even manifestly GI LIV fermion operators of \textquotedblleft
mass\textquotedblright\ type, such as $\overline{\Psi}n_{\lambda}%
\gamma^{\lambda}\Psi$ or $\overline{\Psi}n_{\lambda}\gamma^{\lambda}\gamma
^{5}\Psi$, share this issue. The first can be gauged away by a corresponding
transformation, while the second explicitly breaks GI. This may seem
counter-intuitive, but the problem becomes evident if we calculate the
vector-field polarization loop diagram, since this axial mass term is used for
radiative generation of Cern-Simons term. For example, with $S(k)=1/(\not k%
-\not b\,\gamma^{5})$, one finds
\begin{equation}
p_{\mu}\Pi^{\mu\nu}(p;b)=p_{\mu}\int\frac{d^{4}q}{(2\pi)^{4}}\,\mathrm{Tr}%
[\gamma^{\mu}\,S(q)\,\gamma^{\nu}\,S(q-p)]\neq0
\end{equation}
which explicitly signals a violation of GI \cite{KEP}. \ In principle, an
easier way to check gauge invariance is by examining the modified Compton
scattering matrix element. It is straightforward to see that if the matrix
element takes the form $\xi_{1\mu}(k_{1})\xi_{2\nu}(k_{2})\mathcal{M}^{\mu\nu
}$, where $\xi_{1\mu}(k_{1})$ and $\xi_{2\nu}(k_{2})$ are the photon
polarization vectors, then the gauge invariance condition $k_{1\mu}%
\mathcal{M}^{\mu\nu}=0$ is not satisfied---even at linear order in $b^{\nu}$.

In short, we are demotivated from using GI setups for LIV operators. The
scheme outlined above justifies focusing on corrections that affect the
dispersion relation directly. Moreover, once LIV operators are introduced, the
gauge choice must also be specified, since even a small breaking of GI renders
different gauges inequivalent. For concreteness, in what follows we proceed
with (\ref{LIV-L}) and assume the Standard Model (SM) in the unitary gauge.
While it is interesting to speculate about the form of a LIV setup in the
unbroken electroweak phase that could lead here, for our purposes this is not essential.

With (\ref{LIV-L}) the dispersion relation of the Z boson is modified as
\begin{equation}
Q_{\mu}Q^{\mu}=M_{eff}^{2}=M_{Z}^{2}+\delta_{LIV}Q_{n}^{2} \label{Mass Shell}%
\end{equation}
where $Q_{\mu}$ is the four-momentum of the particle, $Q_{n}\equiv n_{\mu
}Q^{\mu}$ and $M_{Z}$ is the Z boson mass. Here we reintroduce the notion of
an effective mass.

The corresponding decay width can be written as \cite{KEP-JEJ}
\begin{equation}
\Gamma_{eff}(Q)\approx\frac{M_{eff}^{2}}{M_{Z}^{2}}\Gamma_{SM}(Q)=\frac
{M_{eff}^{2}}{Q_{0}M_{Z}}\Gamma_{0SM}%
\end{equation}
where $\Gamma_{SM}$ is SM expression and $\Gamma_{0SM}$ is its rest frame form.

For the propagator we obtain an expression resembling the massive vector
propagator in unitary gauge:%
\begin{equation}
\frac{-i}{Q_{\lambda}^{2}-M_{eff}^{2}}\left(  g_{\mu\nu}-\frac{Q_{\mu}Q_{\nu}%
}{M_{eff}^{2}}\right)  \label{pro0}%
\end{equation}

Usually, unstable massive field propagators are corrected by loop
contributions, which acquire an imaginary part near the pole. By the optical
theorem this imaginary contribution is proportional to the decay rate of the
intermediate particle. Implementing this yields the replacement
\begin{equation}
M_{eff}\rightarrow M_{eff}-iQ_{0}\Gamma_{eff}(Q)/2M_{eff}=M_{eff}%
(1-i\Gamma_{0SM}/2M_{Z})
\end{equation}
which reduces to the often used complex-mass replacement if $\Gamma_{0SM}^{2}$
is dropped. In the Lorentz-invariant limit this reproduces the
well-established propagator \cite{Willenbrock, Greiner-Muller}.

Any further corrections in the numerator of the propagator are suppressed by
$\alpha_{weak}\delta_{LIV}$, where $\alpha_{weak}$ is the weak fine-structure
constant, and are proportional to the $Q_{\mu}Q_{\nu}$ term. In the Drell--Yan
process mediated by the neutral Z boson, which we will analyze in the later
sections, in the limit of massless fermions, all vector and axial currents are
conserved. Terms proportional to momentum in the numerator therefore give no
contribution. Consequently, the working form of the propagator is
\begin{equation}
D_{\mu\nu}=\frac{ig_{\mu\nu}}{Q_{\lambda}^{2}-M_{eff}^{2}(1-i\Gamma
_{0SM}/2M_{Z})^{2}}%
\end{equation}

\section{Phenomenology in Drell--Yan Processes}

The neutral current Drell-Yan process $pp\rightarrow Z/\gamma\rightarrow
\ell^{+}\ell^{-}$ provides the cleanest probe of LIV in the weak sector. This
process is carried by the neutral intermediate bosons: photon, Z boson, and
Higgs, but the Higgs channel is extremely suppressed and therefore its effects
are negligible. In this process, the energy carried by the Z boson is the
highest possible during proton--proton collisions. Consequently, the resonance
region is particularly sensitive: even a small $\delta_{LIV}$ may induce a
measurable shift in the fitted Z mass.

When two protons collide at the LHC, they are arranged so they carry the
following momenta%
\begin{equation}
P_{1}=(E,P\overrightarrow{r})\text{, \ \ }P_{2}=(E,-P\overrightarrow{r})
\end{equation}
where {\normalsize $\overrightarrow{r}$ }is the unit vector along the
collision axis and beam, which is colinear with the detector axis. The high
energy of these protons allows us to neglect the proton mass, and within this
accuracy we can assume that {\normalsize $E=P$. } Partons inside each proton
carry an $x$ portion of the energy, with probability $f_{q_{f}}(x)$. Thus,
when protons collide and the process proceeds via the neutral current, the
momentum carried by the intermediate boson after parton--antiparton
annihilation is:
\begin{equation}
Q=x_{1}P_{1}+x_{2}P_{2}=E((x_{1}+x_{2}),(x_{1}-x_{2})\overrightarrow{r})
\end{equation}
and the cross section of the process has the form
\begin{equation}
\mathrm{\sigma}_{P}=\sum_{f}\int dx_{1}dx_{2}f_{q_{f}}(x_{1})f_{\overline
{q}_{f}}(x_{2})\mathrm{\sigma}_{f}%
\end{equation}
with index $f$ denoting the flavor of the partons inside the proton,
$\mathrm{\sigma}_{P}$ the cross section of the proton--proton collision, and
$\mathrm{\sigma}_{f}$ the parton--antiparton (quark--antiquark) annihilation
cross section. The four--momentum $Q_{\mu}$ is often parametrized by the
invariant mass $M$ and rapidity $Y$:%

\begin{equation}
Q_{\mu}=M(\cosh Y,\overrightarrow{r}\sinh Y)\text{, \ \ }Q_{\mu}^{2}=M^{2}
\label{Q}%
\end{equation}

Here we note that higher rapidity corresponds to higher transferred energy and
3-momentum. If we calculate the Jacobian of this
transformation,{\normalsize \ }%
\begin{equation}
\frac{\partial(M^{2},Y)}{\partial(x_{1},x_{2})}=4E^{2}%
\end{equation}
we can define the differential cross section as%
\begin{equation}
\frac{d^{2}\mathrm{\sigma}_{p}}{dM^{2}dY}=\sum_{f}\frac{f_{qf}(x_{1}%
)f_{\overline{q}f}(x_{2})}{4E^{2}}\mathrm{\sigma}_{f}%
\end{equation}

We will not go into the details of the direct calculation of $\mathrm{\sigma
}_{f}$ ; instead, we cite it from \cite{KEP-JEJ}:%
\begin{align}
\frac{d^{2}\mathrm{\sigma}_{p}}{dM^{2}dY}  &  =\sum_{f}\frac{f_{qf}%
(x_{1})f_{\overline{q}f}(x_{2})}{4E^{2}}\mathrm{\sigma}_{EM}[1+\frac
{g_{q}g_{l}\left(  M^{2}-M_{eff}^{2}(1-\Gamma_{0SM}^{2}/4M_{Z}^{2})\right)
}{2\left\vert e_{f}\right\vert M^{2}\sin^{2}2\theta_{w}}R_{s}\nonumber\\
&  +\frac{(1+g_{q}^{2})(1+g_{l}^{2})}{16e_{f}^{2}\sin^{4}2\theta_{w}}R_{s}]
\label{D cross-section}%
\end{align}
where $e_{f}$ is the charge of corresponding quark, $\mathrm{\sigma}_{EM}$ the
electromagnetic cross section, and $R_{s}$ resonance factor:
\begin{equation}
\mathrm{\sigma}_{EM}=\frac{4\pi\alpha^{2}}{9M^{2}}e_{f}^{2}\text{,
\ \ \ }R_{s}=\frac{M^{4}}{\left(  M^{2}-M_{eff}^{2}(1-\Gamma_{0SM}^{2}%
/4M_{Z}^{2})\right)  ^{2}+M_{eff}^{4}\Gamma_{0SM}^{2}/M_{Z}^{2}}%
\end{equation}

From the cross section we see that the resonance value of the invariant mass
$M_{r}$ is now defined as%
\begin{equation}
M_{r}^{2}=M_{eff}^{2}(1-\Gamma_{0SM}^{2}/4M_{Z}^{2})
\end{equation}
with%
\begin{equation}
M_{eff}^{2}=M_{Z}^{2}+\delta_{LIV}M_{r}^{2}(n_{0}\cosh Y-(\overrightarrow
{n}\cdot\overrightarrow{r})\sinh Y)^{2}%
\end{equation}
Thus we can write%
\begin{equation}
M_{r}^{2}\approx M_{Z}^{2}(1+\delta_{LIV}(n_{0}\cosh Y-(\overrightarrow
{n}\cdot\overrightarrow{r})\sinh Y)^{2})-\Gamma_{0SM}^{2}/4 \label{Mr}%
\end{equation}

The peak value of the cross section changes in the following manner:
\begin{equation}
\mathrm{\sigma}_{f\max}\approx\mathrm{\sigma}_{f\max}^{LI}(1-\frac
{\delta_{LIV}\left(  n\cdot Q_{r}\right)  ^{2}}{M_{Z}^{2}})\approx
\mathrm{\sigma}_{f\max}^{LI}(1-\delta_{LIV}(n_{0}\cosh Y-(\overrightarrow
{n}\cdot\overrightarrow{r})\sinh Y)^{2}) \label{sigmamax}%
\end{equation}

LIV effects depend strongly on the preferred direction $n_{\mu}$ and on
rapidity $Y$. The dependence on rapidity is very strong: effects that are
invisible at small rapidities may become glaring at higher rapidities.

We initially postulated $n_{\mu}$ to be a unit vector, but general violation
patterns do not exclude the lightlike case either, nor is anything in our
assumptions or derivation sensitive to this. Therefore, we can still
distinguish three different cases of LIV: timelike, spacelike, and lightlike.
\begin{align}
\text{Time-like:}{\normalsize \quad}n_{\mu}  &  =(1,\vec{0}),\\
\text{Space-like:}\quad n_{\mu}  &  =(0,\vec{n}),\quad\text{with}\quad\vec
{n}^{2}=1,\\
\text{Light-like:}\quad n_{\mu}  &  =(1,\vec{n}).
\end{align}

For pure timelike violation we obtain:
\begin{align}
M_{r}^{2}  &  \approx M_{Z}^{2}(1+\delta_{LIV}\cosh^{2}Y)-\Gamma_{0SM}^{2}/4\\
\mathrm{\sigma}_{f\max}  &  \approx\mathrm{\sigma}_{f\max}^{LI}(1-\delta
_{LIV}\cosh^{2}Y)
\end{align}
Here dependence on the orientation does not exist, since in the timelike case
no anisotropy appears. The dependence on $Y$ is maximally strong. For timelike
violation, separate observation of high-rapidity cases should be the strategy
for LIV studies. Probably this is a good idea for any LIV case. If we want to
constrain $\delta_{LIV}$ from the accuracy of Z boson mass measurement, using
$\Delta M_{Z}$ (Atlas value), we can estimate:
\begin{equation}
\delta_{LIV}\leq\frac{2\Delta M_{Z}}{M_{Z}\cosh^{2}Y}%
\end{equation}
which for $Y=5,6$, offers $10^{-8}(10^{-9})$.{\normalsize \ \ }

For the spacelike violation case, alongside strong rapidity dependence,
anisotropy also appears. Since $\overrightarrow{n}\cdot\overrightarrow
{r}\equiv\cos\beta,$ with $\beta$ the angle between the preferred direction
and the collision axis, the result depends on Earth's orientation in space and
consequently on sidereal time:
\begin{align}
M_{r}^{2}  &  \approx M_{Z}^{2}(1+\delta_{LIV}\sinh^{2}Y\cos^{2}\beta
)-\Gamma_{0SM}^{2}/4\\
\mathrm{\sigma}_{f\max}  &  \approx\mathrm{\sigma}_{f\max}^{LI}(1-\delta
_{LIV}\sinh^{2}Y\cos^{2}\beta)
\end{align}
Unless, by unfortunate combination, $\cos\beta$ is very small, distinct
oscillations in the cross section at high rapidity should appear with sidereal time.

For the lightlike case $n_{\mu}^{2}=0$. If this case is hard to understand
separately, we can at least look at it as a limiting case of timelike or
spacelike violations. When $n_{0}\gg1$, (\ref{Mr}) and (\ref{sigmamax}) assume
a lighlike form:
\begin{align}
M_{r}^{2}  &  \approx M_{Z}^{2}(1+\delta_{LIV}\left(  \cosh Y-\sinh Y\cos
\beta\right)  ^{2})-\Gamma_{0SM}^{2}/4\\
\mathrm{\sigma}_{f\max}  &  \approx\mathrm{\sigma}_{f\max}^{LI}(1-\delta
_{LIV}\left(  \cosh Y-\sinh Y\cos\beta\right)  ^{2})
\end{align}
Similar to the spacelike violation, the lightlike case also exhibits
anisotropy, though of a different character. It is different enough to be
distinguished from spacelike violation. However, this case is still a kind of
hybrid between timelike and spacelike violations.

The conclusion we can quickly draw here is the following: an almost
exponential dependence on rapidity and modulations by sidereal time should be
the main targets of this kind of LIV study. Dependence on rapidity is the more
universal property, while study of sidereal-time signal modulations may be
restricted by the experimental statistics.

\section{Experimental Strategy}

As we saw in the earlier section, the driving force behind the LIV effects
comes from processes with higher rapidity. While the anisotropy appearing in
the spacelike and lightlike cases will leave its mark on experimental data,
large rapidities remain a prerequisite for LIV detection. If we look at the
standard cross section distribution by rapidity in Drell--Yan
processes\ \cite{Y-distr.}, we can clearly identify that the vast majority of
events occur at small rapidity $Y$, where LIV effects are virtually
nonexistent. Thus, the experimentally acquired data in its vast majority will
appear almost LI, with only a small fraction of events at higher $Y$, where
LIV can in principle become pronounced.

Therefore, to increase the chance of LIV detection we need to isolate the
LIV-sensitive signal by sorting the data according to rapidity and, if
anisotropy is present (in the spacelike and lightlike cases), also by sidereal
time. This allows event selection by transferred momentum and spatial
orientation. Smaller bin sizes for event selection would naturally give a more
accurate picture; however, events with higher rapidities are rare, and there
is possibly a practical limit on bin size.

The pure timelike violation case should be easier to analyze, since there is no need for anisotropy searches. In each rapidity bin, statistics will be
significantly better, and for the cross section near the Z-boson resonance
region we will have slightly different resonance invariant masses and
different peak values. Analysis of the peak's shape, size, and location should
be sufficient to constrain the LIV parameters $\delta_{LIV}$ and $n_{\mu}$.

Let us analyze the timelike violation case as a demonstration of the
above-mentioned strategy. To understand the pure LIV effect, we can plot the
relative difference between $\mathrm{\sigma}_{f}$ and its LI counterpart%

\vspace*{-0.35cm}
{\normalsize \begin{figure}[H]
\caption{\small$(\sigma_{LIV}-\sigma_{LI})/\sigma_{LI}$}%
\label{fig:liv_correction_isolated}%
{\normalsize  \centering
\includegraphics[width=0.85\textwidth]{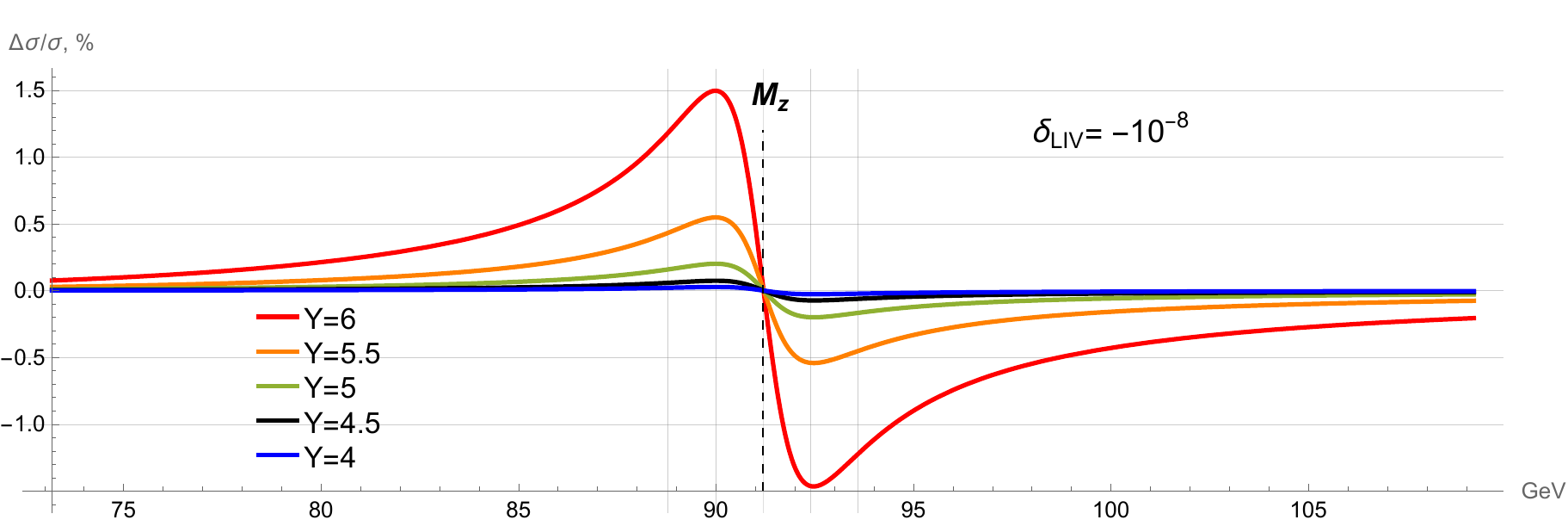}
\includegraphics[width=0.85\textwidth]{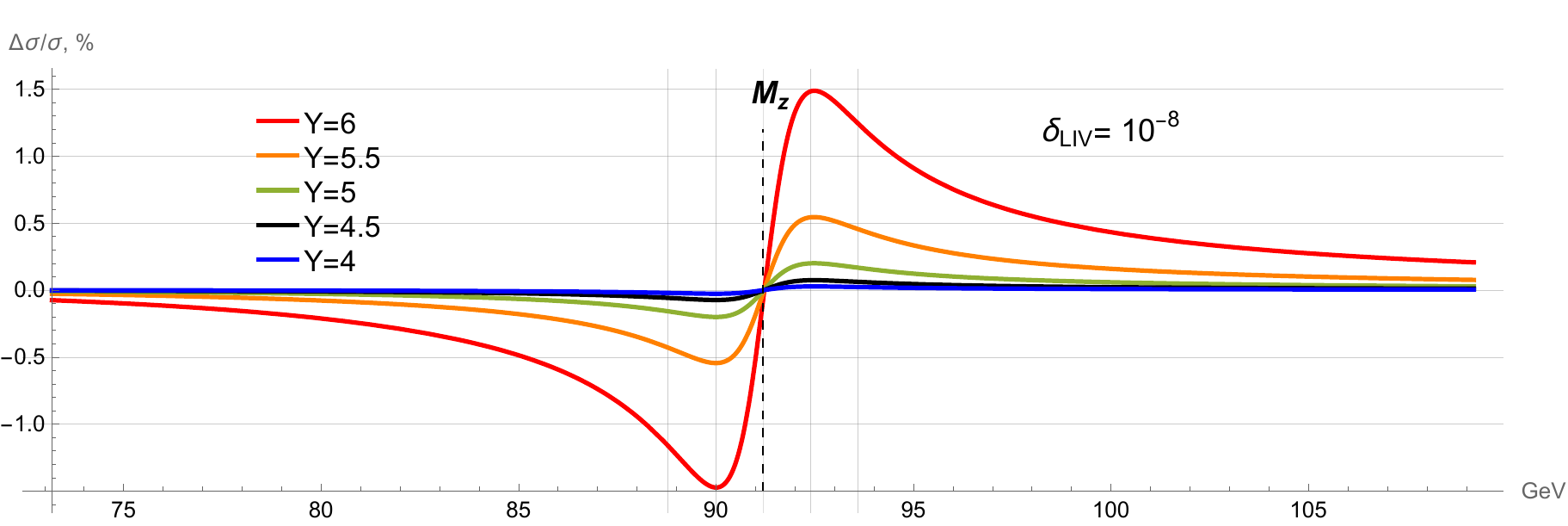}  }
\par
\begin{flushleft}
{\justify {\scriptsize This plot illustrates the behavior of the relative difference between the LIV and LI cross sections at the parton level, thereby isolating the LIV effect for a clearer understanding of its structure. Despite appearances, the LIV effect is not exactly zero at the resonance point. It reaches its maximum value at a point approximately $1.2\ \text{GeV}$ away from the true mass ($M_Z = 91.1876\ \text{GeV}$). }}
\end{flushleft}
\vspace*{-0.5cm}
\end{figure}}

This figure vividly illustrates how the LIV effect is activated near the
resonance mass, and how the enhanced effect quickly dies out farther from the
resonance, scaling as $\delta_{LIV}$ to leading order. Inside the resonance
region at $Y=5,$ the effect is of the order of a few tenths of a percent and
increases up to about 1.5\% at $Y=6$ for $\left\vert \delta_{LIV}\right\vert
=10^{-8}$. The sign of the LIV parameter $\delta_{LIV}$ determines how the
resonance peak shifts, but for both signs the approximate amplitude of the
effect remains the same. This percent-level change is significant but still
too subtle for the human eye to discern on the paper's scale. Therefore, next
we show an exaggerated plot of the LIV and LI cross sections to better
highlight the structure of the LIV effect.%
{\normalsize
%New Vertion of the Figure
}

\begin{figure}[H]
\caption{Highly exaggerated comparison of parton's LIV and LI cross-sections.}%
\label{fig:liv_corrections}%
\centering
\includegraphics[width=1\textwidth]{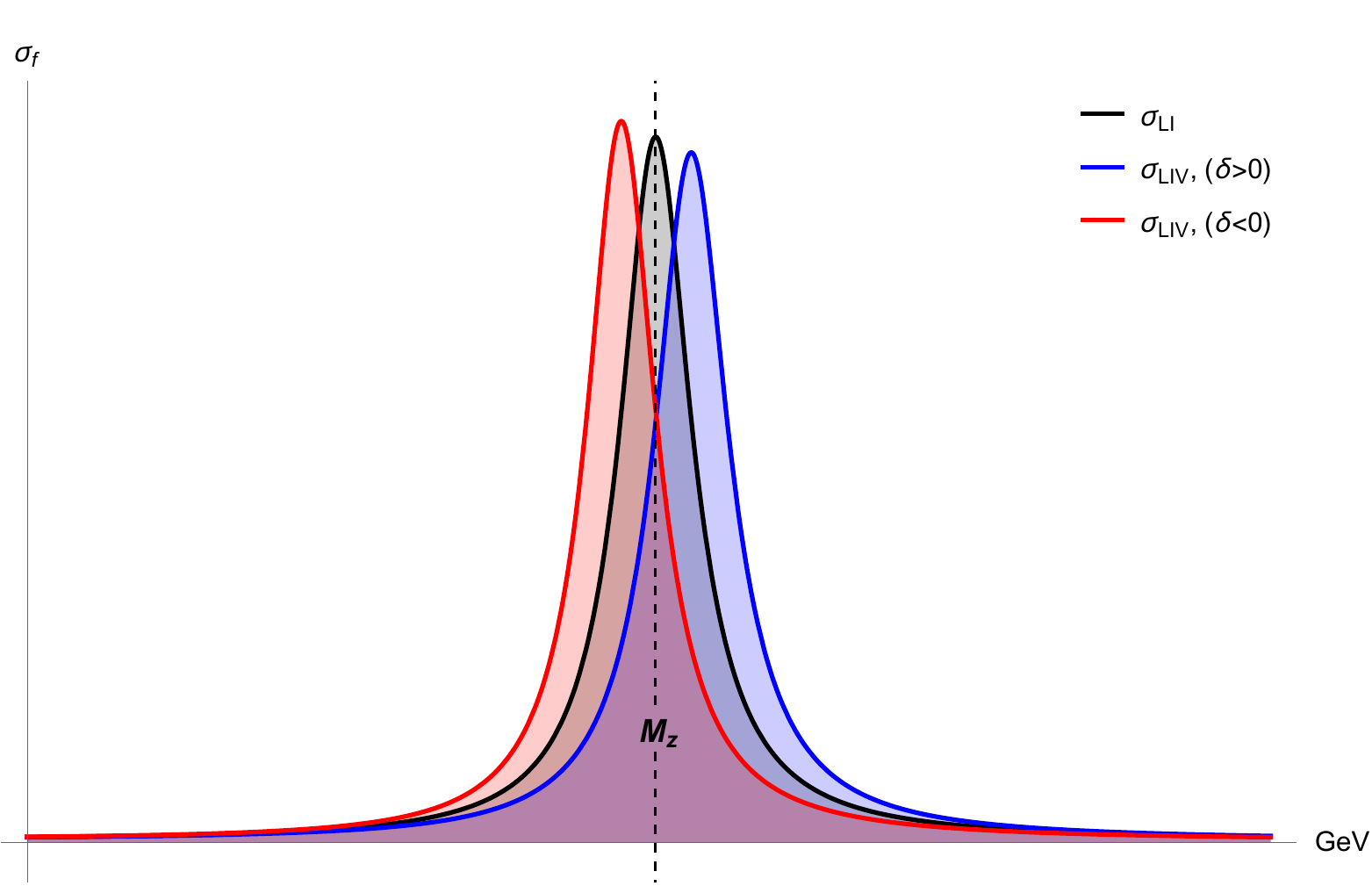}
\includegraphics[width=1\textwidth]{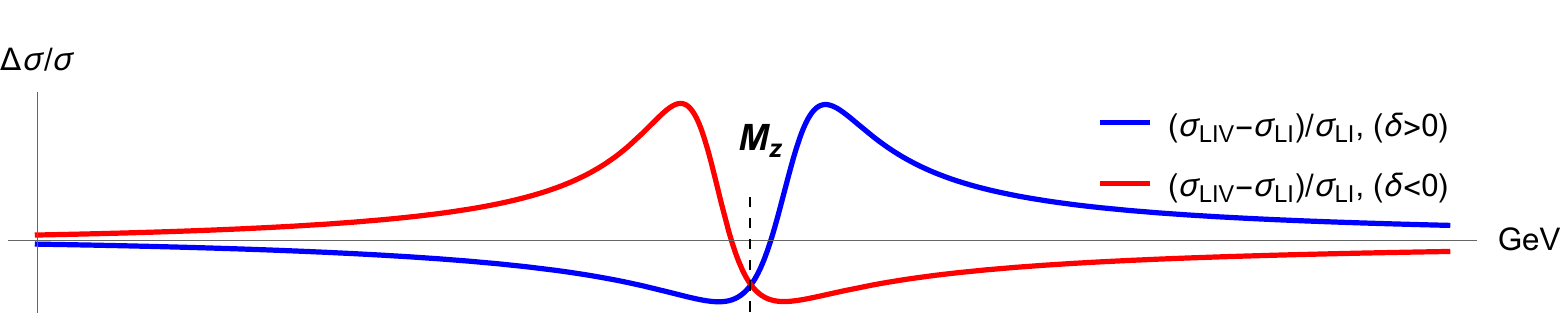}
\par
\begin{flushleft}
{\scriptsize The plot for $Y=8$ illustrates LIV behavior in contrast to LI. Such an explicit presentation is not feasible for $Y=4.5$, where the effect amounts to only $\sim 0.1\%$, making it visually indistinguishable.}
\end{flushleft}
\end{figure}

If LIV is present, the most likely scenario is that all data --- both high
rapidity (where LIV is pronounced) and low rapidity (where LIV is negligible)
--- will be combined together. Attempting to fit this LIV-affected cross
section into an LI template still yields a result, since the effect is
perturbative in nature, but with altered fitting parameters: the extracted
boson mass, and to a lesser degree the decay width. To illustrate this
behavior, below we provide a table of the fitted mass shift $\Delta M_{Z}$
{\normalsize \begin{table}[H]
\caption{Absolute mass shift $|\Delta M_{Z}|$ as a function of rapidity $Y$
and the fractional composition of LI and LIV contributions of cross sections
in the data. \newline}%
\label{tab:z_mass_shift}%
{\normalsize  \centering
\resizebox{\textwidth}{!}{\tiny \setlength{\tabcolsep}{4pt}
\begin{tabular}{>{\centering\arraybackslash}p{1cm}*{11}{>{\raggedleft\arraybackslash}p{1cm}}}
\toprule
${LI}  :$& 100\% & 90\% & 80\% & 70\% & 60\% & 50\% & 40\% & 30\% & 20\% & 10\% & 0\% \\
${LIV} :$& 0\% & 10\% & 20\% & 30\% & 40\% & 50\% & 60\% & 70\% & 80\% & 90\% & 100\% \\
\midrule
Y=0.5 & 0\text{eV} & 0.1\text{keV} & 0.1\text{keV} & 0.2\text{keV} & 0.2\text{keV} & 0.3\text{keV} & 0.3\text{keV} & 0.4\text{keV} & 0.5\text{keV} & 0.5\text{keV} & 0.6\text{keV} \\
Y=1.0 & 0\text{eV} & 0.1\text{keV} & 0.2\text{keV} & 0.3\text{keV} & 0.4\text{keV} & 0.5\text{keV} & 0.7\text{keV} & 0.8\text{keV} & 0.9\text{keV} & 1.0\text{keV} & 1.1\text{keV} \\
Y=1.5 & 0\text{eV} & 0.3\text{keV} & 0.5\text{keV} & 0.8\text{keV} & 1.0\text{keV} & 1.3\text{keV} & 1.5\text{keV} & 1.8\text{keV} & 2.\text{keV} & 2.3\text{keV} & 2.5\text{keV} \\
Y=2.0 & 0\text{eV} & 0.6\text{keV} & 1.3\text{keV} & 1.9\text{keV} & 2.6\text{keV} & 3.2\text{keV} & 3.9\text{keV} & 4.5\text{keV} & 5.2\text{keV} & 5.8\text{keV} & 6.5\text{keV} \\
Y=2.5 & 0\text{eV} & 1.7\text{keV} & 3.4\text{keV} & 5.1\text{keV} & 6.9\text{keV} & 8.6\text{keV} & 10.3\text{keV} & 12.\text{keV} & 13.7\text{keV} & 15.4\text{keV} & 17.1\text{keV} \\
Y=3.0 & 0\text{eV} & 4.6\text{keV} & 9.2\text{keV} & 13.9\text{keV} & 18.5\text{keV} & 23.1\text{keV} & 27.7\text{keV} & 32.4\text{keV} & 37.\text{keV} & 41.6\text{keV} & 46.2\text{keV} \\
Y=3.5 & 0\text{eV} & 12.5\text{keV} & 25.1\text{keV} & 37.6\text{keV} & 50.1\text{keV} & 62.6\text{keV} & 75.2\text{keV} & 87.7\text{keV} & 100.2\text{keV} & 112.7\text{keV} & 125.3\text{keV} \\
Y=4.0 & 0\text{eV} & 34\text{keV} & 68\text{keV} & 102\text{keV} & 136\text{keV} & 170\text{keV} & 204\text{keV} & 238\text{keV} & 272\text{keV} & 306\text{keV} & 340\text{keV} \\
Y=4.5 & 0\text{eV} & 92\text{keV} & 185\text{keV} & 277\text{keV} & 370\text{keV} & 462\text{keV} & 0.6\text{MeV} & 0.6\text{MeV} & 0.7\text{MeV} & 0.8\text{MeV} & 0.9\text{MeV} \\
Y=5.0 & 0\text{eV} & 251\text{keV} & 0.5\text{MeV} & 0.8\text{MeV} & 1.0\text{MeV} & 1.3\text{MeV} & 1.5\text{MeV} & 1.8\text{MeV} & 2.0\text{MeV} & 2.3\text{MeV} & 2.5\text{MeV} \\
Y=5.5 & 0\text{eV} & 0.7\text{MeV} & 1.4\text{MeV} & 2.0\text{MeV} & 2.7\text{MeV} & 3.4\text{MeV} & 4.1\text{MeV} & 4.8\text{MeV} & 5.5\text{MeV} & 6.1\text{MeV} & 6.8\text{MeV} \\
Y=6.0 & 0\text{eV} & 1.9\text{MeV} & 3.7\text{MeV} & 5.6\text{MeV} & 7.4\text{MeV} & 9.3\text{MeV} & 11.1\text{MeV} & 13.0\text{MeV} & 14.8\text{MeV} & 16.7\text{MeV} & 18.6\text{MeV} \\
Y=6.5 & 0\text{eV} & 5\text{MeV} & 10\text{MeV} & 15\text{MeV} & 20\text{MeV} & 25\text{MeV} & 30\text{MeV} & 35\text{MeV} & 40\text{MeV} & 45\text{MeV} & 51\text{MeV} \\
Y=7.0 & 0\text{eV} & 14\text{MeV} & 27\text{MeV} & 41\text{MeV} & 55\text{MeV} & 69\text{MeV} & 82\text{MeV} & 96\text{MeV} & 110\text{MeV} & 124\text{MeV} & 137\text{MeV} \\
\bottomrule
\end{tabular}
}  }
\par
\begin{flushleft}  {\footnotesize {\justify
This table shows how the resonance mass shifts from the true mass value when an LIV-contaminated cross section is reconstructed as an LI Standard Model fit. The greater the contamination by LIV effects---which corresponds to events at higher rapidities---the larger the mass shift. Although real data would consist of a distribution of events across all possible rapidities, this simplified picture still serves as a clear demonstration. Depending on the event selection process, a different pattern may emerge. In particular, selecting only higher-rapidity events would yield a stronger LIV signal in the form of a mass shift. The sign of $\delta_{LIV}$ determines whether the resonance mass is overestimated ($\delta_{LIV} > 0$) or underestimated ($\delta_{LIV} < 0$), but the difference in both cases remains within the displayed accuracy. For this reason, we combine both cases into a single chart. }}

\end{flushleft}
\par
{\normalsize
%\caption{$\delta M_{Z}$ mass shift as a function of rapidity and the fractional composition of Lorentz-invariant and Lorentz-violating contributions in the data}%
}\end{table}}

In the table we see a numerical confirmation of our qualitative expectations.
The low-rapidity cases contain virtually no LIV. If the data is a mixture of
90\% LI and 10\% LIV events at $Y=5$, the total effect is diluted to
$\left\vert \Delta M_{Z}\right\vert \approx0.25$ MeV. By contrast, if 100\% of
$Y=5$ data is analyzed, then $\left\vert \Delta M_{Z}\right\vert \approx2.5$
MeV noticeably larger than the quoted experimental uncertainty of $2.1$ MeV
and therefore impossible to accommodate within the declared accuracy. Clearly,
for $Y=6$ LIV would be even easier to detect, if accelerators could access
such high rapidity regimes.

This table is a kind of proxy intended to mimic the realistic effect of PDFs.
Even in this simplified approximation, the nature of the LIV effect is very
descriptive. Exact calculations using PDFs, or including lower-order
processes, cannot alter the general behavior, though they would certainly
provide a more quantitatively accurate picture.

\section{Discussion: Z vs W Bosons}

We discussed in detail the case of the Z boson because of its narrow width and
clean leptonic channels, properties that make analysis of experimental data
and the chance of discovering possible LIV effects more realistic. We
understand that the modified dispersion relation affects the resonance region
shape in a prominent way, and the effect is more pronounced at higher
rapidities. The sign of the LIV parameter determines whether the shift in
resonance mass is negative or positive; however, the absolute value of the
shift remains approximately the same. Since in all data the pronounced LIV
effect will appear only in a small fraction of events, the total divergence of
the fitted parameters from the Lorentz-invariant ones will be less noticeable.
This warrants dedicated screening of high-rapidity cases in a separate
analysis. In the case of anisotropy, separate binning by sidereal time will be
necessary to understand the nature of the effect, although here we may
encounter the practical limit imposed by insufficient statistics.

While everything said above holds, we must keep in mind that the Z boson is
routinely used for calibration at hadron accelerators (LHC and CDF). This
raises the question of whether such a procedure could bias against potential
LIV effects, and we are not equipped to answer this question yet.

The Z boson looks like an ideal candidate for such a study in a certain sense,
but everything said about the Z boson can in principle be generalized to
charged W bosons as well. Interestingly, recent tensions between Tevatron and
LHC measurements of $M_{W}$, resulting in a discrepancy of about 65 MeV
\cite{MW}, are qualitatively in line with the LIV behavior for negative
$\delta_{LIV}$. It is yet unclear whether this discrepancy originates from
experimental issues, and it is unlikely that the matter will be resolved soon.
Nevertheless, if there is even partial merit to this interpretation, it would
warrant serious exploration of LIV in resonance mass measurements.

Taken together, the Z and W boson cases highlight how collider observables
provide a unique and complementary window on LIV, one that cannot be accessed
through astrophysical probes alone. This motivates the broader conclusions we
now turn to.

\section{Conclusions and Outlook}

In this work we have explored how Lorentz invariance violation (LIV) can
manifest in the weak sector through modifications of the Z-boson dispersion
relation. Starting from a simple but physically motivated Lagrangian
deformation, we showed how only a restricted class of operators leads to
observable effects, with gauge invariance necessarily compromised to ensure
physical LIV. The resulting modifications impact both the propagator and
resonance properties of the Z boson in a calculable way.

The Drell--Yan process provides an especially clean testing ground, as the
resonance region of the Z boson is both experimentally well measured and
theoretically under control. We demonstrated that LIV effects scale almost
exponentially with rapidity and, in anisotropic cases, can introduce sidereal
modulations. This motivates targeted analyses that separate events by rapidity
and, where relevant, by sidereal time. While the majority of experimental data
originates at small rapidities where LIV effects are negligible, the
high-rapidity bins---though rarer---carry the dominant sensitivity. Our
analysis indicates that percent-level modifications of the cross section are
possible for $\left\vert \delta_{LIV}\right\vert $ around $10^{-8}$, leading
to effective shifts in the fitted Z-boson mass that can exceed current
experimental uncertainties.

The timelike violation case offers the clearest starting point, as it avoids
anisotropy and maximizes rapidity dependence, but the spacelike and lightlike
scenarios remain equally important for a comprehensive picture. Our proxy
estimates further show that even after dilution by parton distribution
effects, the characteristic signatures of LIV remain robust.

Beyond the Z boson, similar reasoning extends to W bosons. Recent
discrepancies in W-mass measurements may be qualitatively consistent with
negative $\delta_{LIV}$, although firm conclusions require further scrutiny.
Taken together, the Z and W bosons constitute an essentially unexplored sector
for LIV searches, one that cannot be constrained astrophysically and is
uniquely accessible to collider experiments.

In outlook, we emphasize several directions:

1. Dedicated experimental analyses that implement binning in rapidity and,
where applicable, sidereal time.

2. Extension of the study to W bosons, especially in light of current
experimental tensions.

Altogether, collider studies of unstable bosons provide sensitivity to LIV
possibly up the $10^{-9}$ level, competitive with astrophysical bounds but in
a complementary sector. Pursuing this line of research could therefore open a
new experimental window on fundamental physics beyond the Standard Model.

\section*{Acknowledgments}

 We thank Jon Chkareuli and the participants of the 25th Workshop ‘What Comes Beyond the Standard Models?’ (6–17 July, Bled, Slovenia) for valuable and fruitful discussions, and the organizers for providing a productive working environment. This work was supported by SRNSF (grant STEM‑22‑2604).

\end{document}